\documentclass{phbauth}
\usepackage{epsfig,tabularx}
\begin{document}
\begin{frontmatter}
\title{Dimensional crossover in alternating spin chains}
\author{Adolfo E. Trumper\thanksref{thank1}} and
\author{Claudio J. Gazza}
\address{Instituto de F\'{\i}sica Rosario (CONICET) and
         Universidad Nacional de Rosario, \\
         Boulevard. 27 de febrero 210 bis,(2000) Rosario, Argentina.}
\thanks[thank1]{Corresponding author. Fax: +341-54-4821772\\
E-mail: trumper@ifir.edu.ar, E-mail: gazza@ifir.edu.ar}
\begin{abstract}
The effect of antiferromagnetic interchain coupling in
alternating spin (1,1/2) chains is studied by mean of spin wave
theory and density matrix renormalization group(DMRG). Two limiting
cases are investigated, the two-leg ladder and its two dimensional
(2D) generalization. For the 2D case, spin wave approximation
predicts a smooth dimensional crossover keeping the ground state
ordered, whereas in the ladder case the DMRG results show a
gapped ground state  for any $J_{\perp}>0$. Furthermore, the
behavior of the correlation functions  closely resemble the
uniform spin-1/2 ladder. However, for small $J_{\perp}$,  the gap
behaves quadratically as $\Delta\sim0.6 J^2_{\perp}$. Similarly to
uniform spin chains, it is conjectured an analogous spin gap
behavior for an arbitrary number of mixed spin chains.\\
\end{abstract}
\end{frontmatter}

\begin{figure}
\centerline{\epsfig {figure=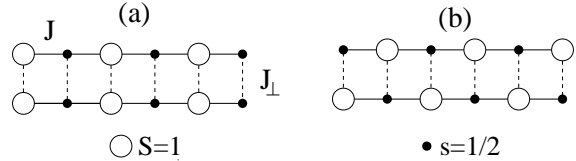,width=75mm,angle=0}} 
\caption{Two alternating spin systems: (a) has a singlet ground state, 
(b) has a ferrimagnetic one.}
\label{2d}
\end{figure}

In the last two decades, quantum magnetism in low dimensions has
become a  main topic in condensed matter physics. Among the
several systems studied in this area, quasi-onedimensional magnets
with mixed spin composition are of special interest since their
recent synthesization\cite{verdaguer,hagiwara}. In particular,
such ferrimagnetic compounds are all composed by weakly coupled
alternating spin chains (Fig. \ref{2d}(b) represents the ladder
case of such topology). There is another family of compounds like
$MnCu(pba)(H_2O)_3.2H_2O$, with pba=1,3-propylenebis(oxamato),
where the ground state is a non-magnetic one, although it is
composed of weakly coupled alternating spin ($Mn^{II}Cu^{II}$)
chains (Fig. \ref{2d}(a)).
This last topology, motivated us to
study  the ground state properties of an alternating spin ladder
(ASL) of type (a) and its 2D generalization with the following
Hamiltonian,
$$
 H=J\sum_{\langle i,j
\rangle_{\|}} {\bf S_i}. {\bf s_j} +J_{\perp} \sum_{\langle
i,j\rangle_{\perp}} \left( {\bf S_i}.{\bf S_j } + {\bf s_i}. {\bf
s_j}\right),
$$
 where ${\bf S_i} ({\bf s_i})$ represents a spin-1
(spin-1/2), $\langle i,j\rangle_{\|}$ ($\langle
i,j\rangle_{\perp}$) denotes nearest neighbors along horizontal
(vertical) direction, with $J$ and $J_{\perp}$ defined positive.

Recently, it was shown how the spin wave (SW) series converges
correctly to the DMRG results in ferrimagnetic
chains\cite{ivanov}. Hence, SW seems to be a good starting point
technique to study the crossover from 1D to 2D, where SW is even
more reliable. 
\begin{table}
\centering
\caption{Spin wave prediction for the relative magnetizations vs.
$J_{\perp}$ for the 2D case.}
\begin{tabular}{p{0.8cm}p{0.8cm}p{0.8cm}p{0.8cm}p{0.8cm}p{0.8cm}p{0.8cm}}
\hline
        $J_{\perp}$   &  0    & 0.2   &  0.4  & 0.6   & 0.8   & 1.0  \\
\hline
         $m_1/S$      & 0.695 & 0.721 & 0.739 & 0.746 & 0.738 & 0.733 \\
\hline
         $m_2/s$      & 0.390 & 0.597 & 0.644 & 0.712 & 0.730 & 0.739 \\
\hline
\end{tabular}
\label{table1}
\end{table}
In table \ref{table1} we show the relative values 
of the magnetizations $m_1/S$ and $m_2/s$, for the spin-1 and spin-1/2, 
respectively, as a function of $J_{\perp}$. It can be 
noticed how the relative values tend to a similar value around 0.73 
as the isotropic limit is reached. In this case we
obtained a value $m_2/s=0.739$ bigger than the uniform spin-1/2 case
$m/s=0.606$\cite{Anderson}. Using mean-field arguments, such a robust 
magnetization is due to the more stronger field caused by  the spin-1 
neighbors.
As general result we observed that the effect of 2D coupling is
the enhancement of  the magnetizations, being the crossover
completely smooth. On the other hand, for the ladder case,
Fig. \ref{2d}(a), it can be proved that quantum fluctuations prevent 
the ordering. This feature was previously pointed out by Fukui and 
Kawakami\cite{fukui} and, in agreement with our DMRG findings, suggests
the opening of a spin gap that is impossible to predict within a
SW framework.

\begin{figure}[htb]
  \centerline{\epsfig {figure=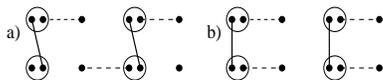,height=1.cm,width=5.0cm,angle=0}}
  \caption{(a) and (b) represent two possible configurations 
   among which the ground state resonates once  $J_{\perp}$ is decreased 
   coming from strong coupling.}
  \label{rung}
\end{figure}

For the two-leg ASL we used the DMRG method\cite{whiteorig}, using 
both, the finite and infinite algorithm, with open boundary conditions 
to study the spin gap and the correlations functions respectively with  
$J_{\perp}$.
Most of the calculations was carried out keeping between 200 and 400 
density matrix states, with a truncation error 
$O(10^{-8})$ at worst and $O(10^{-12})$ in the best case\cite{ourpaper}. 
\begin{figure}
  \centerline{\epsfig{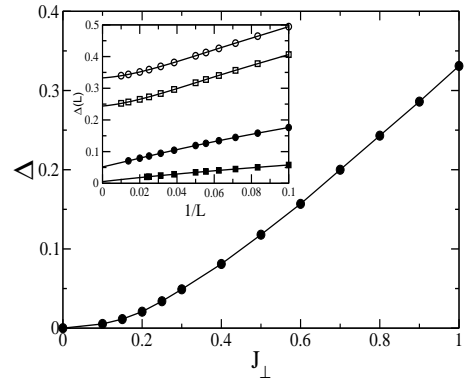}}
  \caption{Spin-gap versus $J_{\perp}$ obtained with DMRG. Inset:
  Spin gaps as a function of $1/L$ for $J_{\perp} = 1$ (empty
  circle), $0.8$ (empty square), $0.3$ (filled circle), and $0.1$
  (filled square). Solid lines are polynomial fits of the
  form\cite{white} $\Delta(L)=\Delta+a_1/L+a_2/L^2+a_3/L^3+....$
  where, $\Delta(L)=E(L,S_z=1)-E(L,S_z=0)$ with $E(L,S_{z})$ the ground
  state energy for a "chain" with $L$ rungs (ranging from 10 to 100), 
  and $z$ component of total spin $S_{z}$.}
  \label{gap}
\end{figure}
In Fig.\ref{gap} we show the gap, that for small values 
of $J_{\perp}$ (below 0.3) behaves quadratically as 
$\Delta\sim 0.6 J^2_{\perp}$ and for
bigger values of $J_{\perp}$ it turns out quite linear
corresponding to the strong coupling regime. For
$J_{\perp}=1$ we obtained $\Delta=0.334$. Consistently with these
features, when extrapolating the gap for different $J_{\perp}$, we
distinguished two scaling regimes at $J_{\perp}\sim 0.3$ as it
is shown in the inset of Fig.\ref{gap}. It is interesting to
compare our results for the weakly coupled regime with that of the
uniform spin-1/2 ladder (USL) that are known to behave linearly
as $\Delta\sim 0.41J_{\perp}$\cite{greven,barnes}. We think that this
discrepancy reflects the quite distinct underlying physics behind
the decoupled regime. In order to get a deeper insight of the
ground state configuration of the ASL we  computed  the local bond
strengths and compared them with the USL. 
\begin{figure}
  \centerline{\epsfig {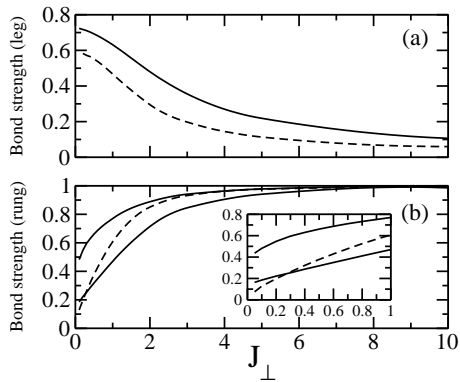}}
  \caption{Normalized bond strength as a function of $J_{\perp}$.
    (a)$\langle {\bf S}_i.{\bf s}_{i+1} \rangle/A$ along the leg for ASL 
       (solid line) and $\langle {\bf s}_i.{\bf s}_{i+1} \rangle/B$ for USL     
       (dashed line), and 
    (b) The upper solid line is for spin-1 rungs $\langle {\bf S}_i.{\bf       
S}_{i+1} \rangle/C$ while the lower solid line is for spin-1/2 rungs $\langle 
{\bf s}_i.{\bf s}_{i+1} \rangle/D$. The constants $A=-1, B=-3/4,$ and $C=-2$ 
are the corresponding free bond cases. The dashed line is the USL case. In the 
inset, the region where the crossing occurs has been amplified.}
  \label{vrcorrel}
\end{figure}
In Fig.\ref{vrcorrel}(a), we present the normalized values of 
$\langle {\bf S}_i.{\bf s}_{i+1} \rangle$ along the leg. A 
similar monotonic behavior with $J_{\perp}$ is observed for ASL and USL.
In Fig.\ref{vrcorrel}(b) we show the normalized values of $\langle
{\bf S}_i.{\bf S}_{i+1}\rangle$ and $\langle {\bf s}_i.{\bf
s}_{i+1} \rangle$, corresponding to the vertical spin-1 and
spin-1/2 rungs, respectively. It can be noticed that the spin-1
rungs remain always larger than the spin-1/2 ones in the ASL. This
is non-trivial at all since the quantities are normalized to its
free bond case. In the strong coupling regime the ground state is
composed by an alternated collection of spin-1 and spin-1/2
singlets. 
As $J_{\perp}$ is decreased, each kind of spin will
resonate among its different neighbors reducing their correlation
along the rungs. Nevertheless, from the valence bond solid picture, 
it is easy to realize (see Fig.\ref{rung}), that the reduction of the
correlation in the spin-1 rung will be less important than in
the spin-1/2 rung. It is also remarkable the
crossing between the spin-1/2 rungs values for ASL and USL that 
occurs at $J_{\perp}\sim 0.3$. Even if we have not a
physical explanation for that  we believe that our results for
the gap and the local bond strengths are connected and indicate a
change of regime for the above value of $J_{\perp}$. 
In addition we studied the spin correlations functions along the 
legs\cite{ourpaper}. We observed that the correlation functions 
along the leg decay exponentially for the whole range of interchain 
coupling, with $\xi^{ASL}\rightarrow\infty$ in the limit
$J_{\perp}\rightarrow 0$, like in USL\cite{greven}. When $J_{\perp}$ 
is strictly zero there is a transition to two ferrimagnetically 
ordered chains in the ASL. Using a semi-log plot of such correlations, 
we estimated the correlation length for $J_{\perp}=0.1$ being 
$\xi^{ASL}\sim 30$, and $\xi^{USL}\sim 25$ lattice spacings. On the 
other hand, for $J_{\perp}=1$ we estimated $\xi^{ASL}\sim5$, whereas 
in the USL it was found $\xi^{USL}\sim 3$ \cite{greven,white}. Except 
for the differences found in the weakly coupled regime, our DMRG
results suggest a strong similarity between ASL and USL.

Finally, from the two limiting cases we have studied, it is
possible to argue a more general statement about the behavior of
the spin gap for an arbitrary number of coupled mixed spin chains.
First, we can rigorously say that any odd number of chains have 
always a ferrimagnetic ground state, due to Lieb-Mattis 
theorem\cite{mattis}, so, it will be gapless --but ordered--.
Then, complementing our results with the conjecture that an even
number of chains will be gapped and that similarly to uniform
ladders\cite{elbiorice,white,rice} this gap will decrease to
zero in the 2D limit, it is recovered an analogous spin gap
behavior to the uniform spin-1/2 case. However, for a finite odd 
number of chains, these gapless states have a completely different 
nature.

This work was done under PICT grant N03-03833 (ANPCYT). The authors 
acknowledge partial financial support from Fundaci\'on Antorchas.

\end{document}